\documentclass[12pt]{article}
\usepackage{amssymb}
\newfont{\twelvemsb}{msbm10 scaled\magstep1}
\newfont{\eightmsb}{msbm8}
\newfam\msbfam
\textfont\msbfam=\twelvemsb
\scriptfont\msbfam=\eightmsb
\catcode`\@=11
\def\Bbb{\ifmmode\let\next\Bbb@\else
  \def\next{\errmessage{Use \string\Bbb\space only in math mode}}\fi\next}
\def\Bbb@#1{{\fam\msbfam{{#1}}}}

\newcommand{\sect}[1]{\setcounter{equation}{0}\section{#1}}

\newcommand{\be}{\begin{equation}}
\newcommand{\ee}{\end{equation}}
\newcommand{\bea}{\begin{eqnarray}}
\newcommand{\eea}{\end{eqnarray}}
\newcommand{\nonu}{\nonumber\\}
\newcommand{\beo}{\begin{eqnarray*}}
\newcommand{\eeo}{\end{eqnarray*}}

\newcommand{\cD}{\mbox{${\cal D}$}}

\newcommand{\cG}{\mbox{${\cal G}$}}
\newcommand{\cU}{\mbox{${\cal U}$}}

\newcommand{\cS}{\mbox{${\cal S}$}}

\newcommand{\cV}{\mbox{${\cal V}$}}
\newcommand{\cW}{\mbox{${\cal W}$}}

\newcommand{\cwgs}{\cW(\cG,\hspace{.5ex}\cS)}

\newcommand{\CC}{\mbox{${\Bbb C}$}}
\newcommand{\II}{\mbox{${\Bbb I}$}}
\newcommand{\JJ}{\mbox{${\Bbb J}$}}

\newcommand{\WW}{\mbox{${\Bbb W}$}}
\newcommand{\ZZ}{\mbox{${\Bbb Z}$}}
\newcommand{\eps}{{\varepsilon}}
\newcommand{\half}{\frac{1}{2}}
\newcommand{\pc}[1]{\mbox{circ. perm. }({#1})}
\newcommand{\mb}[1]{\ \mbox{ #1 }\ }

\newcommand{\PL}[1]{Phys.\ Lett.\ {\bf #1}}

\newcommand{\CMP}[1]{Comm.\ Math.\ Phys.\ {\bf #1}}

\newtheorem{prop}{Proposition}
\newcounter{proposition}
\def\prop#1#2{\refstepcounter{proposition}
  \[\left[
  \parbox[l]{15cm}{
    \begin{minipage}[l]{15cm}
      { \sf $\,$ Proposition \theproposition:} {\rm #1}
      \begin{center}
        \begin{minipage}[c]{14cm}
          {\em #2}
        \end{minipage}
      \end{center}
    \end{minipage}
    }
\right.\]
}

\newtheorem{conj}{Conjecture}
\newcounter{conjecture}

\begin{document}
\newpage
\pagestyle{empty}
\setcounter{page}{0}


\newcommand{\LAP}{LAPTH}
\def\logo{{\bf {\huge LAPTH}}}
\centerline{\logo}

\vspace {.3cm}

\centerline{{\bf{\it\Large 
Laboratoire d'Annecy-le-Vieux de Physique Th\'eorique}}}

\centerline{\rule{12cm}{.42mm}}

\vspace{20mm}
\begin{center}
  {\LARGE  {\sffamily 
    Yangian realisations from finite \cW-algebras}}\\[1cm]

\vspace{10mm}
  
{\large E. Ragoucy$^{a,}$\footnote{ragoucy@lapp.in2p3.fr On leave of absence 
from LAPTH.}}\\[.21cm] 
and\\[.21cm]
{\large P. Sorba$^{b,}$\footnote{sorba@lapp.in2p3.fr.}}\\[.42cm]
 $^{a}$ {\em Theory Division, CERN, CH-1211 Gen\`eve 23 \\[.242cm]
 $^{b}$  Laboratoire de Physique Th{\'e}orique \LAP\footnote{URA 14-36 
    du CNRS, associ{\'e}e {\`a} l'Universit{\'e} de Savoie.}\\[.242cm]
    LAPP, BP 110, F-74941  Annecy-le-Vieux Cedex, France. }
\end{center}
\vfill\vfill

\begin{abstract}
We construct an algebra homomorphism between the Yangian $Y(sl(n))$ 
and the finite \cW-algebras $\cW(sl(np),n.sl(p))$ for any $p$. 
We show how this result can be applied to determine 
properties of the finite dimensional representations of such \cW-algebras.
\end{abstract}

\vfill
\rightline{hep-th/9803243}
\leftline{CERN-TH/98-104}
\rightline{\LAP-672/98}
\rightline{March 1998}

\newpage
\pagestyle{plain}
\setcounter{footnote}{0}

\tableofcontents
\newpage

\sect{Introduction}

\indent

In the year 1985, the mathematical physics literature was enriched
with two new types of symmetries: \cW\ algebras \cite{1} and
Yangians \cite{2}. \cW\ algebras showed up in the context of two
dimensional conformal field theories. They benefited from development
owing in particular to their property to be algebras of constant of
motion for Toda field theories, themselves defined as constrained WZNW
models \cite{3}. Yangians were first considered and defined in
connection with some rational solutions of the quantum Yang-Baxter
equation. Later, their relevance in integrable models with non Abelian
symmetry was remarked \cite{4}. Yangian symmetry has been proved for
the Haldane-Shastry $SU(n)$ quantum spin chains with inverse square
exchange, as well as for the embedding of this model in the
$\hat{SU}(2)_1$ WZNW one; this last approach leads to a new
classification of the states of a conformal field theory in which the
fundamental quasi-particles are the spinons \cite{5} (see also
\cite{6}). Let us also emphasize the Yangian symmetry determined
in the Calogero-Sutherland-Moser models \cite{5,7}. Coming back
to \cW\  algebras, it can be shown that their zero modes provide
algebras with a finite number of generators and which close
polynomially. Such algebras can also be constructed by symplectic
reduction of finite dimensional Lie algebras in the same way usual --or
affine-- \cW\ algebras arise as reduction of affine Lie algebras: they
are called finite \cW\  algebras \cite{8} (FWA). This definition
extends to any algebra which satisfies the above properties of
finiteness and polynomiality \cite{9}. Some properties of such FWA's
have been developed \cite{9}-\cite{10} and in particular a large class of
them can be seen as the commutant, in a generalization of the
enveloping algebra $\cU(\cG)$, of a subalgebra $\hat{\cG}$ of a simple
Lie algebra $\cG$ \cite{11}. This feature fo FWA's has been exploited in
order to get new realizations of a simple Lie algebra $\cG$ once knowing
a $\cG$ differential operator realization. In such a framework,
representations of a FWA are used for the determination of $\cG$
representations. This method has been applied to reformulate the
construction of the unitary, irreducible representations of the
conformal algebra $so(4,2)$ and of its Poincar\'e subalgebra, and
compared it to the usual induced representation technics \cite{12}. It
has also been used for building representations of observable algebras
for systems of two identical particles in $d=1$ and $d=2$ dimensions,
the $\cG$ algebra under consideration being then symplectic ones; in
each case, it has then been possible to relate the anyonic parameter
to the eigenvalues of a \cW-generator \cite{10}.

In this paper, we show that the defining relations of a Yangian
are satisfied for a family of FWA's. In other words, such \cW\ algebras
provide Yangian realizations. This remarkable connection between two a
priori different types of symmetry deserves in our opinion to be
considered more closely. Meanwhile, we will use results on the
representation theory of Yangians and start to adapt them to this class of
FWA's. In particular, we will show on special examples --the algebra
$\cW(sl(4), n.sl(2))$-- how to get the classification of all their
irreducible finite dimensional representations.

It has seemed to us necessary to introduce in some detail the two main
and a priori different algebraic objects needed for the purpose of
this work. Hence, we propose in section \ref{defW} a brief reminder on
\cW-algebras with definitions and properties which will become useful
to establish our main result. In particular, a short paragraph presents
 the Miura transformation. The structure of $\cW(sl(np),n.sl(p))$
algebras is also analysed. Then, the notion of Yangian $Y(\cG)$ is
introduced in section \ref{yang}, with some basic properties on its
representation theory.

Such preliminaries allow us to arrive well-equipped for showing the
main result of our paper, namely that 
{\bf there is an algebra homomorphism between the 
Yangian $Y(sl(n))$ and the
  finite $\cW(sl(np),n.sl(p))$ algebra (for any $p$).}
This property is proven in section \ref{YWcl} for the classical case
(i.e. \cW-algebras with Poisson brackets) and generalized to the
quantum case (i.e. \cW-algebras with usual commutators) in the section
\ref{YWq}. The proof necessitates the explicit knowledge of
commutation relations among \cW\ generators. Such a result is
obtained in the classical case via the soldering procedure
\cite{sold}. Its extension to the quantum case leads to determine
$sl(n)$ invariant tensors with well determined symmetries. In order
not to overload the paper, all these necessary intermediate results
are gathered in the appendices. Finally, as an application, the
representation theory of $\cW(sl(2n),n.sl(2))$ algebras is considered
in section \ref{Wrepres}. General remarks and a discussion about some further
possible developments conclude our study.

\sect{Finite \cW\  algebras: notation and classification.\label{defW}}

\indent

As mentioned above, the \cW\  algebras that we will be interested in can
be systematically obtained by the Hamiltonian reduction technique in a
way analogous to the one used for the construction and classification
of affine \cW\  algebras \cite{3,13,14}. Actually, given a
simple Lie algebra $\cG$, there is a one-to-one correspondence between
the finite \cW\  algebras one can construct in $\cU(\cG)$ and the $sl(2)$
subalgebras in $\cG$. We note that  
any $sl(2)$  \cG-subalgebra is
principal in a subalgebra $\cS$ of
$\cG$. The step generator $E_+$ in the $sl(2)$ subalgebra
which is principal in $\cS$ is then written as a linear combination of
the simple root generators of $\cS$: $E_+=\sum_{i=1}^s\, E_{\beta_i}$ where 
$\beta_i$, $i=1,\dots,s=\mbox{rank}\cS$ are the simple roots of \cS. It can be 
shown that one can complete uniquely $E_+$ with two generators $E_-$ and $H$ 
such that $(E_\pm,H)$ is an $sl(2)$ algebra. It is rather usual to
denote the 
corresponding \cW\ 
algebra as \cwgs. It is an algebra freely generated by a
finite number of 
generators and which has a second antisymmetric product. Depending
on the  nature
of this second product, we will speak of classical (the product is a
Poisson bracket) 
or a quantum (the product is a commutator) \cW-algebra.

\subsection{Classical \cwgs\ algebras}

\indent

To specify the Poisson structure of the \cW-algebra, we start with the 
Poisson-Kirilov structure on $\cG^*$. It mimicks the Lie algebra
structure  on \cG, 
and we will still denote by $H_i,\ E_{\pm\alpha}$ the generators in $\cG^*$.
$\cU(\cG^*)$ is then a Poisson-Lie enveloping algebra. We construct the 
classical \cwgs\ algebra from an Hamiltonian reduction on $\cU(\cG^*)$, 
the constraints being given by the $sl(2)$ embedding as follows.

The Cartan generator $H$ of the $sl(2)$ subalgebra under consideration
provides 
a gradation of \cG:
\be
\cG =\oplus_{i=-N}^{N}\cG_i \mbox{ with } [H,X]=i\,X,\ \forall\,
X\in\cG_i
\ee
The root system of \cG\ is also graded:
$\Delta=\oplus_i\Delta_i$. We have
\be
H\in\cG_0,\ 
E_\pm\in\cG_{\pm1}
\mbox{ and }E_-=\sum_{\alpha\in\Delta_{-1}}\chi_\alpha\, E_\alpha,
\ \chi_\alpha\in\CC.
\ee
Then, the first class constraints are
\begin{equation}
\Phi_\alpha=\Phi(E_\alpha)=E_\alpha=0\ \mbox{ if }\ E_\alpha\in\cG_{<-1}
\ \mbox{ and }\ \Phi_\alpha=\Phi(E_\alpha)=E_\alpha-\chi_\alpha=0\  
\mbox{ if }\ 
\ E_\alpha\in\cG_{-1}
\end{equation}
 the second class constraints being given by
\begin{equation}
\Phi(X)=X=0\ \mbox{ if }\ X\in\cG_{\geq0}
\ \mbox{ and }\ \{E_+,X\}\neq0
\end{equation}
Note that considering $\cG^*$ as a module of the above mentionned $sl(2)$,
the generators of \cwgs\ are in one-to-one correspondence with
the highest weights of this $sl(2)$.

The Poisson bracket structure of the \cwgs\ algebra is then given by the 
Dirac brackets associated to the constraints. More explicitely,  one labels 
the constraints $\Phi_i,\ i=1,\dots,I_0$ and denotes them by 
$C_{ij}\sim\{\Phi_i,\Phi_j\}$ where the symbols $\sim$ means that one 
has to apply the 
constraints after calculation of the Poisson-Kirilov brackets. Then, 
the Dirac brackets are given by
\begin{equation}
\{A,B\}_*\ \sim\
\{A,B\}-\sum_{i,j=1}^{I_0}\{A,\Phi_i\}\, C^{ij}\,\{\Phi_j,B\}
\ \mbox{ with }\ C^{ij}=(C^{-1})_{ij}
\end{equation}
They have the remarquable property that any generator in $\cU(\cG^*)$ has 
vanishing Dirac brackets with any constraint, so that the \cwgs\ algebra
can be seen as the quotient of $\cU(\cG^*)$ provided with the Dirac bracket by 
the ideal generated by the constraints.

\subsection{Quantum \cwgs\ algebras}

\indent

These algebras are a quantization of the classical \cwgs\ algebras, where
the Poisson structure has been replaced by a commutator. 
As we have an algebra on which one has imposed some constraints, there
are two ways to quantise it: either, one first quantize $\cU(\cG^*)$
and then impose the constraints at the quantum level; or one
quantize directly the \cW-algebra. In the first case, 
 one can just look at \cU(\cG) as the quantization of
$\cU(\cG^*)$ and then use a BRS
formalism. This
has been developped in \cite{dBT}, where the BRS operator is
constructed, and its cohomology computed. The quantum version of the 
\cwgs\ algebra is then the zeroth cohomological space, which
inherits an algebraic structure from \cU(\cG). 
It is possible to
explicitely construct a representative of each cohomology classes, using
the highest weights in \cG\ of the $sl(2)$ under consideration. Note
that for super \cW-algebras, the same treatment can be applied, the
factorisation of spin $\half$ fields (and other gauging properties)
being replaced by a filtration of the cohomological spaces in that
formalism \cite{MR}.

Here, we will look directly at the
quantization of the \cW-algebra. 
In that approach, we start with the classical \cwgs\ algebra and
ask for an algebra which has a non commutative product law, the
associated commutator admitting as a limit the Poisson bracket (PB). 
This more pedestrian
and less powerful approach will be sufficient for our purpose. In
fact, we will only need to know that the quantum \cwgs\ algebra
has the same number of generators as the classical one, as well as 
the same leading term in the 
commutator. By leading term, we mean
that in each commutator, the term of highest ``conformal spin'' is the
same as the right hand side of the corresponding PB. In the following,
we will choose this term to be a symmetrized product (see below).

\subsection{Miura representations\label{miura}}

\indent

Associated to the gradation of \cG, there exists a representation of
the \cwgs-algebra, called the 
Miura map. It is an algebra morphism from $\cU(\cG^*_0)$ to \cwgs. 
Note that one can show that $\cU(\cG^*_0)$ has the same dimension as 
\cwgs, 
but the Miura map is neither an algebra isomorphism, nor a vector 
space isomorphism. It is based on a restriction of the Hamiltonian
reduction that leads from \cG\ to \cwgs.
More explicitely, for classical finite \cW-algebras, 
we start with the ($\cG^*$-valued) matrix
\be
J_0=e_-+j^k\, t_k \mb{with} [h,t_k]=0
\ee
where $e_-$ and $t_k$ are in the fundamental representation of
\cG\ ($\{t_k\}$ is a basis of $\cG_0$) and $j^k\in\cG^*_0$. Then, we 
consider the transformations
\be
J_0\ \rightarrow\ J^g=g\, J_0\, g^{-1} \mb{with} g\in G_+,\ \
\mbox{Lie}(G_+)=\cG_+ 
\ee
It can be shown that there exists a unique element $g$ such that
\be
J_0\ \rightarrow\ J^g=J_{hw}=e_-+W^k\, m_k \mb{with} [e_+,m_k]=0
\label{eq.miura}
\ee
where $m_k$ (in the fundamental representation of
\cG) are the highest weights of $(h,e_\pm)$, the $sl(2)$
algebra under consideration.  $\{W^k\}$ generate the classical
\cW-algebra and are determined by the Miura map (\ref{eq.miura}). They
are expressed in terms of the $j^k$ generators, and thus this map
allows to construct the \cwgs-algebra generators as polynomials in
$\cG_0$ ones (see e.g. \cite{3} for details).

At the quantum level, one can still proceed in two ways: either work
on cohomological space (the Miura map in that case corresponds to a
restriction to the zero grade of the general cohomological
construction), 
or directly quantize the classical Miura construction.

Using the Miura construction then leads to (finite dimensional) 
representations of
\cwgs\ associated to (finite dimensional) representations of $\cG_0$. 
Note however
that the irreducibility of these \cwgs-representations is a priori not known,
even if one starts from an irreducible representation of $\cG_0$ (see
\cite{these} for a counter example).

\subsection{Example: $\cW(sl(np),n.sl(p)))$ algebras\label{exemple}}

\indent

As an example, let us first consider the $\cW(sl(4), sl(2) \oplus sl(2))$
algebra.

It is made of seven generators $J_i,\ S_i\ (i=1,2,3)$ and a central
element $C_2$ such that:
\begin{equation}
\begin{array}{ll}
  {[}J_i, J_j] = i{\epsilon_{ij}}^k J_k & i,j,k\, =\, 1,2,3  \\
  {[} J_i, S_j ] = i{\epsilon _{ij}}^k S_k &  \\
  {[} S_i, S_j] = i{\epsilon _{ij}}^k J_k\, (C_2 -2\ {\vec{J}\,}^2)
\ \ & \\
  {[}C_2, J_i ] = [C_2 , S_i] =0 & \mbox{with }\ {\vec{J}\,}^2 =
  J^2_1 + J^2_2 + J_3^2
\end{array}
  \label{1}
\end{equation}
We recognize the $sl(2)$ subalgebra generated by the $J_i$'s as well
as a vector representation (i.e. $S_i$ generators) of this $sl(2)$
algebra. We note that the $S_i$'s close polynomially on the other generators. 

\indent

The same type of
structure can be remarked, at a higher level, for the class of
algebras $\cW(sl(np),n.sl(p))$ where $n.sl(p)$ stands for the regular
embedding  $sl(p)
\oplus \cdots \oplus sl(p)$ ($n$ times). We recall that an algebra $\cS$ 
is said to be regular in $\cG$, itself generated
by the Cartan generators $H_i,\ i=1, \dots,$ rank $(\cG)$ and the root
generators $E_\alpha ,\ \alpha \in \Delta$, if $\cS$ is, up to
a conjugation, generated by a subset of the above Cartan part, as well
as a subset $\{E_\alpha \},\ \alpha \in \delta  \subset \Delta $ of the
$\cG$-root generator set. In order to determine the
number and the ``conformal spin'' of the \cW\  generators, one has first
to decompose the adjoint representation of $\cG$ under the $sl(2)$ which is
principal in $\cS=n.sl(p)$: $\cG= \oplus_j \cD_j$ with $\cD_j$ the
$(2j+1)$-dimensional irreducible representation of $sl(2)$. 
Then to each $\cD_j$ will correspond
one \cW\  generator of ``conformal spin'' ($j+1$) if we keep in mind
that such a generator can be seen as the zero mode of a primary field
in a (Toda) conformal field theory. Actually, it is known
 that the $\cG$ adjoint
representation can be seen as arising from the direct product of the
$\cG$ fundamental representation by itself. 
Since this later representation reduces with respect to the $\cS$
principal $sl(2)$ as $n\cD_{\frac{p-1}{2}}$, it leads to\footnote{For
  details about the conformal spin contents of \cW-algebras computed using
  $sl(2)$ representations see \cite{14}.}:
\begin{equation}
n\cD_{\frac{p-1}{2}} \times n\cD_{\frac{p-1}{2}} \longrightarrow n^2
(\cD_{p-1} + \cD_{p-2} + \cdots + \cD_1) + (n^2 -1) \cD_0
  \label{2}
\end{equation}

A more careful study will allow to recognize in the $(n^2-1)\cD_0$ the
 generators of an $sl(n)$ algebra, and to associate to each
set of $(n^2-1)\cD_k,\ k=1,2,\dots,p-1$, an irreducible (adjoint)
representation under the above determined $sl(n)$ algebra, the
corresponding elements being of ``conformal spin'' $(k+1)$. 
To each remnant $\cD_k$
will finally be associated a spin $(k+1)$ element which commutes with
all the \cW\  generators: these central elements will be 
denoted $C_2, C_3, \dots, C_p$. They can be identified with the first
$p$ Casimir operators of the $sl(np)$ algebra.

\indent

Using a notation which will become clear in the next sections, we
call $W^a_0,\ a=1, \dots, n^2-1$ the $sl(n)$ generators, and $W^a_k,
\ k= 1,2,\dots, p-1$, the \cW\  generators of respective spin $(k+1)$. We
can gather the above assertions in the following commutation relations:
\begin{equation}
\begin{array}{ll}
  {[}W_0^a, W_0^b ] = {f^{ab}}_c W_0^c & a=1,2, \dots, n^2-1 \\
  {[}W^a_0, W^b_k ] = {f^{ab}}_c W^c_k & k=1,2,\dots, p-1 \\
  {[}C_i, W^a_0 ] = [C_i, W^a_k] =0 \ \ \ \ \ \ & i=2,3,\dots, p 
\end{array}
  \label{3}
\end{equation}
The remaining commutator takes the form
\be
{[}W^a_k, W^b_\ell ] = P^{ab}_{k\ell}(W)
\ee
where $P^{ab}_{k\ell}$ is a polynomial in the \cW-generators, which is
of ``conformal spin'' $(k+\ell-1)$. It is determined using the
technics described above.

The $\cG_0$ subalgebra associated to $\cW(sl(np),n.sl(p))$ is 
$p.sl(n)\oplus (p-1).gl(1)$ which one can denote $s(p.gl(n))$, 
i.e. traceless matrices of 
$gl(n)\oplus gl(n)\oplus\cdots\oplus gl(n)$ ($p$ times).
Thus, the Miura map for this kind of \cW-algebras leads to  
a realisation in term of 
generators of the enveloping algebra of $s(p.gl(n))$.

\sect{Yangians $Y(\cG)$\label{yang}}
\subsection{Definition}
We briefly recall some definitions about Yangians, most of them being
gathered in \cite{CP}.
Yangians are one of the two well-known families of infinite dimensional 
quantum groups 
(the other one being quantum affine algebras) that correspond to 
deformation of the
universal enveloping algebra of some finite-dimensional Lie algebra, 
called \cG. 
As such, it is a Hopf algebra, topologically generated by elements $Q_0^a$ and 
$Q_1^a$, $a=1,\dots,\mbox{dim}\cG$ which satisfy the following 
defining relations:
\begin{eqnarray}
&& Q_0^a \mbox{ generate } \cG\ :\ {[Q_0^a,Q_0^b]}={f^{ab}}_c\, Q_0^c 
\label{algG}\\
&& Q_1^a \mbox{ form an adjoint rep. of } \cG\ :\ {[Q_0^a,Q_1^b]}=
{f^{ab}}_c\, Q_1^c \\
&& {[Q_1^a,[Q_0^b,Q_1^c]]} + {[Q_1^b,[Q_0^c,Q_1^a]]} 
+ {[Q_1^c,[Q_0^a,Q_1^b]]} =
{f^a}_{pd}{f^b}_{qx}{f^c}_{ry}{f^{xy}}_e
\eta^{de}\ s_3(Q_0^p,Q_0^q,Q_0^r)\ \ \  \label{yG}\\
&& {[[Q_1^a,Q_1^b],[Q_0^c,Q_1^d]]} + 
 {[[Q_1^c,Q_1^d],[Q_0^a,Q_1^b]]}\ =\ \nonu
&&\ \ \ \left(
{f^a}_{pe}{f^b}_{qx}{f^{cd}}_y{f^y}_{rz}{f^{xz}}_g+
{f^c}_{pe}{f^d}_{qx}{f^{ab}}_y{f^y}_{rz}{f^{xz}}_g\right)
\eta^{eg}\ s_3(Q_0^p,Q_0^q,Q_1^r) \label{ysl2}
\end{eqnarray}
where ${f^{ab}}_c$ are the totally antisymmetric structure constants of \cG, 
$\eta^{ab}$ is the Killing form, and $s_n(.,.,\dots,.)$ is the totally
symmetrized product of $n$ terms.
The generators $Q^a_n$ for $n>1$ are defined recursively through
\be
{f^{a}}_{bc}\, [Q_1^b,Q_{n-1}^c]\ =\ c_v\, Q_{n}^a
\mb{ with }
c_v\eta^{ab}={f^{a}}_{cd}{f^{bcd}}
\label{Qn-1}
\ee
It can be shown that for $\cG=sl(2)$, (\ref{yG}) is a consequence of the other 
relations, while for $\cG\neq sl(2)$, (\ref{ysl2}) follows from 
(\ref{algG}--\ref{yG}). The coproduct on $Y(\cG)$ is given by 
\begin{equation}
\begin{array}{l}
\Delta(Q_0^a)=1\otimes Q_0^a +Q_0^a\otimes1\\
\Delta(Q_1^a)=1\otimes Q_1^a +Q_1^a\otimes1 +\half{f^a}_{bc}\ Q_0^b\otimes
Q_0^c
\end{array}\label{eq:7}
\end{equation}
In the following, we will focus on the Yangians $Y(sl(n))$.

\subsection{Evaluation representations of $Y(sl(n))$\label{secteval}}

\indent

When $\cG=sl(n)$, there is a special class of finite dimensional irreducible 
representations called evaluation representations. They are 
defined from the algebra homomorphisms
\begin{equation}
ev_A^\pm\ \left\{
\begin{array}{ccc}
Y(sl(n)) & \rightarrow & \cU(sl(n)) \\
Q_0^a & \rightarrow & t^a \\
Q_1^a & \rightarrow & A\, t^a \pm\, {d^a}_{bc}\, t^b t^c
\end{array}\right.\ \mbox{ with }
A\in\CC
\end{equation}
where the $t^a$'s form a $sl(n)$ basis, and ${d^a}_{bc}$ 
is the totally symmetric invariant tensor of 
$sl(n)$ (we set  
${d^a}_{bc}=0$ when $n=2$). It can be shown that $ev_A^+$ and $ev_A^-$ are 
isomorphic (and indeed $ev_A^+=ev_A^-=ev_A$ when $\cG=sl(2)$). 

An evaluation representation of $Y(\cG)$ is defined by the pull-back
of a \cG-representation (with the help of the evalutation
homomorphism $ev_A^\pm$). The corresponding representation space will 
be denoted
generically by $\cV_A^\pm(\pi)$ where $\pi$ is a representation
of $sl(n)$. 
We select hereafter two properties \cite{CP} which will
be used in section \ref{Wrepres}.

\indent

{\bf Theorem 1:} 
{\it Any finite-dimensional irreducible $Y(sl(n))$ module is 
isomorphic to a subquotient of a tensor product of evaluation 
representations.}

\indent

{\bf Theorem 2:} 
{\it When \cG=sl(2), let $\cV_A(j)$ be the $(2j+1)$-dimensional
irreducible representation space of $ev_A$ ($j\in\half\ZZ$). Then, 
$\cV_A(j)\otimes\cV_B(k)$ is reducible if and only if $A-B=\pm(j+k-m+1)$
for some $0<m\leq\mbox{min}(2j,2k)$.

In that case, $\cV_A(j)\otimes\cV_B(k)$ is not completely reducible, and 
not isomorphic to $\cV_B(k)\otimes\cV_A(j)$;
otherwise, $\cV_A(j)\otimes\cV_B(k)$ is irreducible and isomorphic to
$\cV_B(k)\otimes\cV_A(j)$.}

\sect{Yangians and classical \cW-algebra\label{YWcl}}

In this section, we want to show that there is an algebra morphism 
between the Yangian $Y(sl(n))$ and the
classical $\cW(sl(np),n.sl(p))$ algebras ($\forall\ p$). For such a 
purpose, we need to compute some of the PB of the \cW-algebra. It is done 
using the soldering procedure \cite{sold}, the calculation
being quite tricky (see appendices). For the generic case of 
$\cW(sl(np),n.sl(p))$ algebras, 
the result is
\begin{eqnarray}
\{W_1^a,W_1^b\} &=& \frac{1}{5}{f^{ab}}_c\,
 W_2^c-\frac{p^3}{16}\,\left(
{d^{au}}_{v}{f^{b}}_{cu}-{d^{bu}}_{v}{f^{a}}_{cu} 
\right){d^{v}}_{de}\, W_0^cW_0^dW_0^e\label{pbw1}
\end{eqnarray}
where the indices run from $0$ to $n^2-1$, with the notation
$W_0^0=0$ and $W_1^0=C_{2}$ and the normalisation:
\be
\{W_0^a,W_k^b\} = \frac{1}{p}{f^{ab}}_c\, W_k^c\ \ \ \ k=0,1,2,...
\ee
When $p=2$, we have the constraint $W_2^a=5\,{d^a}_{bc}W_0^bW_1^c$.
Let us stress that the tensors ${f^{ab}}_c$ and ${d^{ab}}_{c}$ are
$gl(n)$ tensors, \underline{not} $sl(n)$ ones: see appendix \ref{gln} for
clarification. 

\indent

{F}or the case of $\cW(sl(2p),2.sl(p))$, the relations simplify to:
\begin{eqnarray}
\{W_1^a,W_1^b\} &=&{f^{ab}}_c\,\left[
\frac{1}{5}\, W_2^c-\frac{p^3}{2}\, 
W_0^c\,{\vec{W_0}}^2\right]\\
\{W_1^a,W_2^b\} &=& {f^{ab}}_c\ \left[
\frac{3}{14}\, W_3^c+ \frac{6(3p^2-7)}{p(p^2-1)}\, W_1^cW_1^0
+\frac{(p^2-9)(p^2-4)}{2(p^2-1)}\ W_1^c\,{\vec{W_0}}^2 
+\right. \nonu
&&\left.\phantom{{f^{ab}}_c\ }+3\,W_2^0W_0^c
-30\,W_0^c\,(\vec{W_1}\cdot\vec{W_0})\rule{0mm}{2.76ex}\right]\label{pbw2}
\end{eqnarray}
together with the constraints 
$W_2^a=10\,[\,W_0^aW_1^0+\delta^a_0(\vec{W_1}\cdot\vec{W_0})\,]$ for
$p=2$, and $W_3^a=0$ for $p=2$ or 3.

In this basis, the map is
\begin{equation}
\rho_p\ \left\{
\begin{array}{cccc}
Y(sl(n)) & \rightarrow & \cW(sl(np),\,n.sl(p)) &  \\ \\
Q_k^a & \rightarrow & \beta_k\, W_k^a &  \mbox{ for }k=0,1,\dots,p \\ \\
Q_{p+l}^a & \rightarrow & P^a_{p+l}(W_0,W_1,\dots,W_p) &  \mbox{ for }l>0 
\end{array}\right.\label{rhop}
\end{equation}
where $P^a_l$ are some homogeneous polynomials which preserve the
``conformal spin'' of $W_k^a$. A careful computation shows (see
appendix \ref{gln}), 
using the PBs (\ref{pbw1}--\ref{pbw2}),
 that the generators $W_k^a$ obey the relations 
(\ref{algG}--\ref{ysl2}), the commutators being replaced by PBs. 
As $Y(sl(n))$ is topologically
generated by $Q_0^a$ and $Q_1^a$, it is sufficient to give 
$\beta_0$ and $\beta_1$. 
Indeed, once (\ref{rhop}) is satisfied for $k=0$ and 1, the relation
(\ref{Qn-1}) together with the PB of the \cW-algebra ensure that 
(\ref{rhop}) can be iteratively constructed for all $k$.
We show in the appendix \ref{gln} that 
in our basis this relation is indeed satisfied for
\be
\beta_0=p \mb{and}\beta_1=2\label{normclass}
\ee
We can thus conclude:
\prop{}{
The classical algebra $\cW(sl(np),n.sl(p))$ provides a representation of the
Yangian $Y(sl(n))$, the map being given by $\rho_p$ defined in
(\ref{rhop}) and (\ref{normclass}).
}

\sect{Yangians and quantum $\cW(sl(np),n.sl(p))$ algebras
\label{YWq}}

We can use the above study to deduce the same result for quantum 
$\cW(sl(np),n.sl(p))$ algebras. In fact, as these algebras are a 
quantisation of 
the classical ones, we can deduce that the most general form of the 
commutator is
\begin{eqnarray}
{[W_0^a,W_k^b]} &=& \frac{1}{p}{f^{ab}}_c\, W_k^c \\
{[W_1^a,W_1^b]} &=& {f^{ab}}_c\,
\frac{1}{5}\, W_2^c-\frac{p^3}{16}
\left({d^{au}}_{v}{f^{b}}_{cu}-{d^{bu}}_{v}{f^{a}}_{cu} 
\right){d^{v}}_{de}\, s_3(W_0^c,W_0^d,W_0^e)+\nonu
&&+t^{ab}_c\  W_1^c +t^{ab}_{cd}\ s_2(W_0^c,W_0^d)+
\tilde{t}^{ab}_c\  W_0^c
\eea
for $sl(n)$, and in the special case of $sl(2)$
\bea
{[W_1^a,W_2^b]} &=& {f^{ab}}_c\left[
\frac{3}{14}\ W_3^c+\frac{6(3p^2-7)}{p(p^2-1)}\, s_2(W_1^c,W_1^0)
+3\,s_2(W_2^0,W_0^c)+\right.\nonu
&& \left.\phantom{{f^{ab}}_c\ }+\left(
\frac{(p^2-9)(p^2-4)}{2(p^2-1)}\,\eta_{dg}\eta^c_{e}-30\,\eta_{de}\eta^c_{g}
\right)\ 
s_3(W_0^g,W_0^d, W_1^e)\right]
+\nonu
&&
+g^{ab}_c\  W_2^c +g^{ab}_{cd}\ W_1^cW_0^d+
g^{ab}_{cde}\ s_3(W_0^c,W_0^d,W_0^e) 
+\hat{g}^{ab}_c\  W_1^c 
+\nonu
&&+\hat{g}^{ab}_{cd}\ s_2(W_0^c,W_0^d)+
\tilde{g}^{ab}_c\ W_0^c
\end{eqnarray}
for some tensors $t^{ab}_{a_1a_2\cdots a_k}$ 
and $g^{ab}_{a_1a_2\cdots a_k}$. 
By construction, the $t$-tensors are symmetric in the lower indices
and antisymmetric in the upper ones. The $g$-tensors are only
symmetric in the lower indices, except for $g^{ab}_{cd}$ which has no
symmetry property.
Moreover, the Jacobi identities with $W_0^a$ show that they are 
invariant tensors.
Hence, we are looking for objects which belong to the trivial representation
in $\Lambda_2(\cG)\otimes S_k(\cG)$, $S_2(\cG)\otimes S_k(\cG)$, or
$\Lambda_2(\cG)\otimes \Lambda_2(\cG)$, where 
$S_k(\cG)$ is the totally symmetric product 
$\cG\otimes\cG\otimes\cdots\otimes\cG$
($k$ times), while $\Lambda_2(\cG)$ is the antisymmetric product 
$\cG\otimes\cG$. 
Computing the decomposition of these tensor products shows that 
the multiplicity $M_0$
of the trivial representation in these products is (for $\cG=sl(n)$):
\be
\begin{array}{cc}
M_0[\Lambda_2(\cG)\otimes \cG] = 1 & M_0[S_2(\cG)\otimes \cG]=
\left\{\begin{array}{l}1\mbox{ if }n\neq2\\
0 \mbox{ for }sl(2)\end{array}\right.
\\ & \\
M_0[\Lambda_2(\cG)\otimes S_2(\cG)]=\left\{\begin{array}{l}1
\mbox{ if }n\neq2\\
0 \mbox{ for }sl(2)\end{array}\right. 
 &
M_0[\Lambda_2(\cG)\otimes \Lambda_2(\cG)] = \left\{\begin{array}{l}3
\mbox{ if }n\neq2\\
1 \mbox{ for }sl(2)\end{array}\right. \\ & \\
M_0[\Lambda_2(\cG)\otimes S_3(\cG)] = \left\{\begin{array}{l}4
\mbox{ if }n\neq2,3\\
3 \mbox{ for }sl(3)\\
1 \mbox{ for }sl(2)\end{array}\right. &
\end{array}
\label{mult}
\ee
Now, it is easy to show that the following tensors indeed belongs to 
these spaces\footnote{More general formulae are given in
  the appendix \ref{tensor}.}:
\begin{equation}
\begin{array}{ll}
\Lambda_2(\cG)\otimes \cG\ :\ \ \  t^{ab}_{c} = {f^{ab}}_{c} &
S_2(\cG)\otimes \cG\ :\ \ \ t^{ab}_{c} = {d^{ab}}_{c} \\ \\
\Lambda_2(\cG)\otimes S_2(\cG)\ :\ \ \ 
t^{ab}_{cd} = {f^{ab}}_{e}{d^{e}}_{cd}&
\end{array}
\end{equation}

As they are evidently independent and give the correct 
multiplicities (with the convention that the $d$-tensor is null 
for $sl(2)$), we deduce that
 the most general form one gets is:
\begin{eqnarray}
{[W_1^a,W_1^b]} &=& -\frac{p^3}{16}
\left({d^{au}}_{v}{f^{b}}_{cv}-{d^{bu}}_{v}{f^{a}}_{cv} 
\right){d^{v}}_{de}\, s_3(W_0^c,W_0^d,W_0^e)+\nonu
&&+{f^{ab}}_c\, \left[ 
\frac{1}{5}\, W_2^c+\mu_1\, W_1^c +\mu_2\, {d^c}_{de}\ s_2(W_0^d,W_0^e)+
\mu_3\,  W_0^c\right]\label{comw1}
\end{eqnarray}
for the algebra $\cW(sl(np),n.sl(p))$. This commutator is the only one
needed to prove that the algebra satisfies the defining relations of
the Yangian when $n\neq2$.

{F}or the algebra $\cW(sl(2p),2.sl(p))$, we need also the relation:
\begin{eqnarray}
{[W_1^a,W_2^b]} &=&{f^{ab}}_c\left[
\frac{3}{14}\ W_3^c+\frac{6(3p^2-7)}{p(p^2-1)}\, s_2(W_1^c,W_1^0)
+3\,s_2(W_2^0,W_0^c)\right.
+\nonu
&&\phantom{{f^{ab}}_c\ }+
\left(\frac{(p^2-9)(p^2-4)}{2(p^2-1)}\,\eta_{dg}\eta^c_{e}
-30\,\eta_{de}\eta^c_{g} \right)\ 
s_3(W_0^g,W_0^d, W_1^e)+\label{comw2}\\
&&\phantom{{f^{ab}}_c\ }+\nu_1\, W_2^c +\nu_1'\, W_1^c
+\nu_1''\,W_0^c
+\nu_2\,{f^{c}}_{de}\, W_1^dW_0^e+ 
\left.\nu_3\,
\eta_{de}\, s_3(W_0^c,W_0^d,W_0^e)\rule{0mm}{2.76ex}\right]\nonumber
\end{eqnarray}

Then, one can show that the commutators (\ref{comw1}--\ref{comw2}) 
obeys to the defining 
relations of the Yangians $Y(sl(n))$ for the same normalisations as in
the classical case. It is done in the appendices \ref{apWq} and \ref{tensor}.
\prop{}{
The quantum algebra $\cW(sl(np),n.sl(p))$ provide a representation of the
Yangian $Y(sl(n))$, the map being given by $\rho_p$ defined in
(\ref{rhop}) and (\ref{normclass}).
}
At this point, let us note that the Yangian structure of the algebra
$\cW(sl(4),2sl(2))$ has been already remarked \cite{9,Ge} and used for
quantum mechanics applications \cite{Ge}.

\sect{Representations of $\cW(sl(2n),\, n.sl(2))$ algebras\label{Wrepres}}

Owing to the above identification, it is possible to adapt some known
properties on Yangian representation theory to finite $\cW$
representations. We first illustrate this assertion in the case of
$\cW(sl(2n),n.sl(2))$.

\indent

\prop{}{Any finite dimensional irreducible representation of 
the algebra
$\cW(sl(4), 2.sl(2))$ is either an evaluation module $\cV_A(j)$ 
or the tensor product
of two evaluation modules $\cV_A(j)\otimes\cV_{(-A)}(k)$.
\newline
Conversely, $\cV_A(j)$ for any $A$, and $\cV_A(j)\otimes\cV_{(-A)}(k)$
($A \neq0$) 
with $2A \neq \pm (j+k -m+1)$ for any $m$
such that $0<m\leq$ min $(2j,2k)$, are 
finite dimensional irreducible representations of 
the algebra
$\cW(sl(4), 2.sl(2))$. 
\newline
The tensor product is calculated
via the Yangian coproduct defined in (\ref{eq:7}).}

\indent

The proof is done by direct calculation, using the theorem 2 of
section \ref{secteval}. As a (irreducible) representation of the 
$\cW(sl(4), 2.sl(2))$ algebra must be a (irreducible) representation
of the Yangian $Y(sl(2))$, we deduce that the (finite dimensional) 
irreducible
representations of $\cW(sl(4), 2.sl(2))$ are in the set of evaluation
modules  $\cV_A(j)$ or 
$\cV_A(j)\otimes\cV_{B}(k)\otimes\cdots\otimes\cV_{C}(\ell)$.

For $\cV_A(j)$, it is obvious that we have an irreducible
representation, where the
value of the \cW\ Casimir operator $C_2$ is related to $A$:
\be
C_2(A,j)\ =\ \left(2j(j+1)+A^2\right)\,\II
\ee
For a product $\cV_A(j)\otimes\cV_{B}(k)$, calculations show that we
must have 
\be
A+B=0 \mb{ and } 
C_2(A,j\,;\,B,k)\ =\ \left(2j(j+1)+2k(k+1)+\half(A^2+B^2)\right)
\,\II\otimes\II
\ee
in order to get a representation of the \cW-algebra.
The irreducibility is fixed by the first part of theorem 2, section
\ref{secteval}, i.e. when there is no $m\in\,]\,0,\min(2j,2k)\,]$ such that
$2A=\pm(j+k-m+1)$. 

It is the second part of the theorem which ensures that the above
construction exhausts the set of irreducible finite dimensional
representations of $\cW(sl(4),2sl(2))$. Indeed, in the product 
$\cV_A(j)\otimes\cV_{B}(k)\otimes\cV_C(\ell)$, we already know that we
must have $B=-A$ and $C=-B$ for $\cV_A(j)\otimes\cV_{B}(k)$ and 
$\cV_{B}(k)\otimes\cV_C(\ell)$ to be representations of the
\cW-algebra. 
Then, the irreducibility of 
$\cV_{(-A)}(k)\otimes\cV_A(\ell)$ implies that this last representation is
isomorphic to $\cV_A(\ell)\otimes\cV_{(-A)}(k)$. Therefore, 
$\cV_A(j)\otimes\cV_{(-A)}(k)\otimes\cV_A(\ell)$ is isomorphic to
$\cV_A(j)\otimes\cV_A(\ell)\otimes\cV_{(-A)}(k)$. But the product
$\cV_A(j)\otimes\cV_A(\ell)$ is not a representation of the
\cW-algebra\footnote{In fact, for $A=0$, the tensor product indeed provides a
  representation of the \cW-algebra (it is just a representation of
  $sl(2)$). However, in that case, the tensor product is not irreducible.}, so that the triple product is not either.

\indent

Note that we get the surprising result that the tensor product of two
representations of the $\cW(sl(4), 2.sl(2))$ algebra ($\cV_A(j)$ and 
$\cV_{B}(k)$) is
\underline{not} always a representation of this algebra. In some
sense, this result can be interpreted as a no-go theorem for the
existence of a coproduct for \cW-algebras. 

Let us also remark that the above representations are those obtained
through the Miura map (see sections \ref{miura} and \ref{exemple}), 
so that we have proved that the Miura map gives
all the irreducible finite dimensional representations of this
\cW-algebra. Moreover, as the $\cG_0$ algebra we have to consider is
just $sl(2)\oplus sl(2)\oplus gl(1)= s(2.gl(2))$, the condition 
$A+B=0$ in the tensor product $\cV_A(j)\otimes\cV_{B}(k)$ can just be
interpreted as the traceless condition on $s(2.gl(2))$. Indeed, a
representation of $2gl(2)$ is given by a representation space
$\cD_j\otimes\cD_k$ of $2sl(2)$, together with the values $A$ and $B$
of the two $gl(1)$ generators, while for $s(2.gl(2))$, one has to
impose $A+B=0$.
In fact, we are able to prove a more general result:

\indent

\prop{}{Any finite dimensional irreducible representation of 
the algebra
$\cW(sl(2n), n.sl(2))$ must be either an evaluation module $\cV^{\pm}_A(\pi)$ 
or the tensor product of two evaluation modules
$\cV^{\pm}_A(\pi)\otimes\cV^{\mp}_{(-A)}({\pi'})$, where 
$\pi$ and $\pi'$ are irreducible finite dimensional representations of 
$sl(n)$, the tensor product being calculated
via the Yangian coproduct defined in (\ref{eq:7}).
\newline
All these representations can be obtained from the Miura
map:
\centerline{$s(2.gl(n))\equiv 2.sl(n)\oplus gl(1)\ \rightarrow\ 
\cW(sl(2n), n.sl(2))$}}

\indent

Note that these algebras are just the ones used in \cite{11} to construct the
finite \cW-algebras  as commutants in $\cU(\cG)$. 

It seems rather
natural to conjecture that this situation will remain valid in the general
case of $\cW(sl(np), n.sl(p))$ algebras \cite{zaugg}.

\sect{Conclusion}

A rather surprising connection between Yangians and finite
\cW-algebras has been developped in this paper. We
have proved directly that finite \cW-algebras of the type $\cW(sl(np),\,
n.sl(p))$ satisfy the defining relations of the Yangian $Y(sl(n))$. In
particular, we have been led to explicitly compute rather non trivial
commutators of \cW\ generators, namely spin 2 - spin 2 and spin 2 -
spin 3 ones, a result which is interesting in itself. The question is
now to understand more deeply this relationship between Yangian and
finite \cW-algebras.

Of course, the structure of the $\cW(sl(np),\, n.sl(p))$ algebra (see
section \ref{exemple}) reveals the special role played by its (spin one) 
Lie subalgebra $sl(n)$. The \cW\ generators of equal spin gather into
adjoint representations of this $sl(n)$ algebra, inducing some
resemblance with the $Y(sl(n))$ yangian structure. 

At this point, let
us remark another common point between $Y(sl(n))$ and $\cW(sl(np),\,
n.sl(p))$, namely the construction of their finite dimensional
representations with the help of $sl(n)$ ones. Indeed, the evaluation
homomorphism (in the case of Yangians) and the Miura map (for
\cW-algebras) play identical roles for such a construction: the former
allows to represent $Y(sl(n))$ on the tensor product of $sl(n)$
representations (with the use of additional constant numbers), while the
later uses a representation of the $\cG_0$ algebra $p.sl(n)\oplus
(p-1).gl(1)$. This clearly shows a one-to-one correspondence.

Let us also stress another feature of the $\cW(sl(np),\, n.sl(p))$
algebras: for $p=2$, they are the commutant in (a localisation of)
$\cU(sl(n))$ of an Abelian subalgebra $\tilde{\cG}$ of $sl(n)$ 
\cite{11,these}, the case $p>2$ being with no doubt generalisable.

Finally, in the seek of understanding our results, one could think to
a R-matrix approach. This point of view looks natural, since a
R-matrix definition of the Yangians is available, while our
\cW-algebras are symmetry algebras of (integrable) non-Abelian lattice
Toda models.

\indent

Due to the wide class of \cwgs-algebras, it seems natural to think of
generalisations of our work. First of all, one could imagine to study
Yangians $Y(\cG)$ (with $\cG\neq sl(n)$) from the \cW-algebras point
of view. However, a rapid survey of \cwgs-algebras shows that 
$\cW(sl(np),\, n.sl(p))$
algebras are the only \cwgs-algebras where the generators are all 
gathered in adjoint
representations of the Lie \cW-subalgebra. Inversely, \cwgs-algebras
might be a way to generalize the notion of Yangians $Y(\cG)$ to cases where the
generators are in any representation of \cG. In that case, the Hopf structure
remains to be determined. Finally, it would be of some interest to
look for an extension to affine \cW-algberas.

\indent

Let us end with two comments concerning applications. The first one
concerns the representation theory of finite \cW-algebras. Preliminary
results have been given in section \ref{Wrepres} and deals with the
classification of finite dimensional representations of
$\cW(sl(2n),\,n.sl(2))$. More complete results will be available soon
\cite{zaugg}. Secondly, the possibility of carrying out the tensor product
of \cW\ representations, although only in special cases, allows to
imagine the construction of spin chain models based on a finite
$\cW(sl(np),\, n.sl(p))$ algebra.

\section*{Acknowledgements}

\indent

We have benefited of valuable discussions with M.L. Ge, 
Ph. Roche and particulary Ph. Zaugg.

\appendix
\section*{Appendices}
\sect{The soldering procedure\label{sold}}

The soldering procedure \cite{sold} allows to compute the Poisson brackets of the
\cW-algebras. 
The basic idea is to implement the \cwgs-transformations from \cG\
ones. Indeed, as the \cwgs-algebra can be realised from an Hamiltonian
reduction on \cG, one can see the \cW\ transformations as a particular
class of (field dependent) \cG\ conjugations that preserve the
constraints we have imposed. Thus, the soldering procedure just says 
that the PBs of
the \cwgs\ algebra can be deduced from the commutators in \cG.
It applies to any \cwgs\ algebra, but we will
focus on the $\cW(sl(np),n.sl(p))$ ones.

For such a purpose,
we define 
\be
J=\sum_{a=1}^{(np)^2-1} J^a\, t_a \mbox{ where } t_a \mbox{ are $(np)\times
  (np)$ matrices and } J^a\in\cG^*
\ee
Then, we introduce the highest weight basis for the $sl(2)$ under
consideration $(E_\pm,H)$:
\be
\JJ=E_-+\sum_{i=1}^{p(n^2)-1} W^i\, M_i \mbox{ with } [E_+,M_i]=0
\label{highw}
\ee
where $M_i$ are $(np)\times (np)$ matrices and 
$E_+$ is considered here in the fundamental representation
$E_+=\sum_{i=1}^{np-1} E_{i,i+1}$, with $E_{ij}$ the matrix whose
elements are $(E_{ij})_{kl}=\delta_{ik}\delta_{jl}$.

To compute the PB of the generator $W^i$ of the \cW-algebra, one
writes the variation of $\JJ$ under the infinitesimal action of one of
the \cW-generators in two ways, namely:
\be
\delta_\eps \JJ = \{\, tr(\eps \JJ),\JJ\, \}_{PB}=[\, \eps \JJ,\JJ\, ]
\ee
where $\{\, tr(\eps \JJ),\JJ\, \}_{PB}$ is the matrix of PB:
\be
\{\, tr(\eps \JJ),\JJ\, \}_{PB}=\{\, tr(\eps \JJ),W^i\, \}_{PB}\, M_i
\ee
and $[\, \eps \JJ,\JJ\, ]$ is a commutator of $(np)\times(np)$ matrices:
\be
{[\, \eps \JJ,\JJ\, ]}\ =\ {f^a}_{bc}\, \eps^b\JJ^c\ t_a
\ee
$\eps$ is an  $np\times np$ matrix such that 
$\delta_\eps \JJ=[\, \eps \JJ,\JJ\, ]$ keeps the form (\ref{highw}) 
with of course 
$\delta_\eps E_-=0$. This matrix $\eps$ has $n(p^2)-1$ free entries, which is the
right number of parameters needed to describe a gauge transformation
by a general element in the \cW-algebra. Identifying the matrix of PB
with the commutator of matrices leads to the PB of the \cW-algebra.

\indent

We now use the property $gl(np)\sim gl(n)\otimes gl(p)$ to explicitely
compute some of the PBs. In $gl(n)\otimes gl(p)$, a general element
can be written as 
\be
J=\sum_{\alpha=0}^{n^2-1} \sum_{s=0}^{p^2-1}J^{\alpha s}\ 
t_\alpha\otimes \tau_s
\ee
with $t_\alpha$, $n\times n$ matrices and $\tau_s$, $p\times p$ matrices.
The principal $sl(2)$ in $n.sl(p)$ takes the form
\be
H=\II_n\otimes h \mbox{ and }E_\pm=\II_n\otimes e_\pm
\ee
where $(h,e_\pm)$ form the principal $sl(2)$ in $sl(p)$, and $\II_n$
is the identity in $sl(n)$. Then,
\be
\JJ=\II_n\otimes e_--\sum_{k=0}^{p-1}  \WW_k\otimes m_k
\ee
where $e_-$ is viewed as a $p\times p$ matrix, 
$e_- = \sum_{i=1}^{p-1} E_{i,i+1}$ and 
$m_k$ are $p\times p$ matrices representing the highest weights 
of the principal $sl(2)$ in $sl(p)$. They have been computed in
\cite{fold}:
\be
m_k=\sum_{i=1}^{p-k}a_k^i\, E_{i,i+k}
\mbox{ with } a_k^i= \frac{(i+k-1)!\, (p-i)!}{(i-1)!\, (p-k-i)!}
\ee
$\WW_k$ are $n\times n$ matrices whose entries $W_k^a$
$(a=0,\dots,n^2-1)$ are the \cW-generators (with 
$W_k^0$ related to $C_{k+1}$ for $k>0$ and $W_0^0=0$ by the 
traceless condition on
$\JJ$). Note that the indices run from 0 to $n^2-1$ because we are
using $gl(n)$ indices instead of $sl(n)$ ones (see appendix \ref{gln} for
precisions).

Using this notation and demanding that $\delta_\eps \JJ$ keeps 
the form (\ref{highw}), one can compute the commutator 
$[\, \eps \JJ,\JJ\, ]$ to get the relations defining the matrix
$\eps$.
These relations are quite awful, but, for our purpose, we just 
need to compute the matrix elements $[\, \eps \JJ,\JJ\, ]_{1,2}$ and 
$[\, \eps \JJ,\JJ\, ]_{1,3}$. A rather long calculation leads to:
\begin{eqnarray}
\{tr(\mu\, \WW_0),\WW_k\}_{PB} &=&  \frac{1}{p} [\WW_k,\mu]\ \ \ k=0,1,2,...
\label{sold1}\\
\{tr(\lambda\, \WW_1),\WW_1\}_{PB} &=& \frac{6}{p(p^2-1)} \left(
\frac{p^2-4}{5}[\WW_2,\lambda]+ 
\frac{1}{2}\left( \rule{0mm}{2.1ex}
[\WW_0,\{\WW_1,\lambda\}]+\{\WW_1,[\WW_0,\lambda]\}\right)\right.+\nonu
&&\left. \phantom{p(p^2-1)\ }
-\half [\WW_0,[\WW_0,[\WW_0,\lambda]]]\right) \\
\{tr(\lambda\, \WW_1),\WW_2\}_{PB} &=& \frac{6}{p(p^2-1)} \left(
\frac{3(p^2-9)}{14}\, [\WW_3,\lambda]
+\{\WW_2,[\WW_0,\lambda]\}+
\half [\WW_0,\{\WW_2,\lambda\}]+\right. \nonu
&& \phantom{p(p^2-1)\ }+\frac{1}{3}[\{\WW_1,\WW_1\},\lambda]-
\half[\WW_1,[\WW_0,[\WW_0,\lambda]]]+\nonu
&& \phantom{p(p^2-1)\ }-
\frac{1}{4}[\WW_0,[\WW_1,[\WW_0,\lambda]]]+ 
\left.-\frac{1}{12}[\WW_0,[\WW_0,[\WW_1,\lambda]]]\rule{0mm}{2.1ex}\right)
\label{sold2}
\end{eqnarray}
where $\mu$ (resp. $\lambda$) is a $n\times n$ matrix whose entries 
$\mu^a$ (resp. $\lambda^a$) are the parameters 
of the infinitesimal transformations associated to $W_0^a$
(resp. $W_1^a$): 
\be
\mu=\mu^a\,t_a\ ;\ \lambda=\lambda^a\,t_a\ ;\ 
\WW_k=W_k^a\, t_a\ \ \ k=0,1,2\ \ \ \mb{with} t_0=\II_n
\label{norm7}
\ee

\sect{Classical $\cW(sl(np),n.sl(p))$ algebras\label{gln}}
\subsection{Generalities}

As we are using heavily the isomorphism $gl(np)\sim gl(n)\otimes
gl(p)$ for our calculations, we are forced to make use of $gl(n)$ indices
instead of $sl(n)$ ones. We denote the last index by $a=0$. It
corresponds to the $gl(1)$ generator that commutes with $sl(n)$ in
$gl(n)$. We can consistently
extend the definition of the totally (anti-)symmetric tensors $f$ and 
$d$ from $sl(n)$ to $gl(n)$ by
\be
{d^{ab}}_0=2\, \eta^{ab}
\mbox{ and } {f^{ab}}_0=0\ \ \forall\ a,b=0,1,\dots,n^2-1
\ee
In the fundamental representation of $gl(n)$, we have then the
decomposition:
\be
t^a\, t^b = \half({f^{ab}}_c+{d^{ab}}_c)t^c
\mb{with}t^0=\II_n
\ee
Then, it is easy to show that the Jacobi identities
\bea
&& {f^{ab}}_c\, {f^{cd}}_e\, +{f^{bd}}_c\, {f^{ca}}_e\, 
+{f^{da}}_c\, {f^{cb}}_e\, =0 \label{Jacoff}\\
&& {d^{ab}}_c\, {f^{cd}}_e\, +{d^{bd}}_c\, {f^{ca}}_e\, 
+{d^{da}}_c\, {f^{cb}}_e\, =0 \label{Jacofd}
\eea
are still valid for any values of $a,b,d,e=0,1,\dots,n^2-1$. 
If we compute
$\{\{t^a,t^b\},t^c\}-\{\{t^c,t^b\},t^a\}=[[t^a,t^c],t^b]$, 
we get also the relation between $f$ and $d$ tensors:
\be
{d^{ab}}_d\, {d^{dc}}_e\, -{d^{bc}}_d\, {d^{da}}_e\, 
={f^{ac}}_d\, {f^{db}}_e \label{dd=f}
\ee
These identities will be the only one needed for our purpose.
Note that the identity
\be
{f^{ab}}_c\, f_{abd}\, =c_v\,\eta_{cd}
\ee
is \underline{not} valid in $gl(n)$ since the left hand side is 0 for
$c=d=0$. 

As an aside comment, let us remark that the isomorphism $gl(np)\sim
gl(n)\otimes gl(p)$ together with the above conventions allow us to
construct the structure constants of $gl(np)$ from those of $gl(n)$
and $gl(p)$. Indeed let $t^a$ (resp. $\bar{t}^q$ and
$T^{(a,q)}=t^a\otimes\bar{t}^q$) be the generators in the fundamental
representation of $gl(n)$ (resp. $gl(p)$ and $gl(np)$); let ${f^{ab}}_c$
 (resp. ${{\bar{f}^{qr}}}_s$ and ${F^{(a,q)(b,r)}}_{(c,s)}$) be
their structure constants; and let ${d^{ab}}_c$
(resp. ${{\bar{d}^{qr}}}_s$ and ${D^{(a,q)(b,r)}}_{(c,s)}$) be their totally
symmetric invariant tensor. The calculation of $[T^{(a,q)},T^{(b,r)}]$
and $\{T^{(a,q)},T^{(b,r)}\}$ show that
\bea
{F^{(a,q)(b,r)}}_{(c,s)} &=& \half\left({f^{ab}}_c\,{{\bar{d}^{qr}}}_s+ 
{d^{ab}}_c\,{{\bar{f}^{qr}}}_s\right)\\
{D^{(a,q)(b,r)}}_{(c,s)} &=&\half\left({f^{ab}}_c\,{{\bar{f}^{qr}}}_s+ 
{d^{ab}}_c\,{{\bar{d}^{qr}}}_s\right)
\eea
which shows that e.g. 
\be
{D^{(a,q)(b,r)}}_{(0,0)}=2{\eta^{(a,q)(b,r)}}=2\eta^{ab}\eta^{qr}
\ee
in agreement with our conventions. 

\indent

With these conventions and properties, we deduce 
from the soldering procedure result (\ref{sold1}-\ref{sold2}) 
the PBs 
\begin{eqnarray}
\{W_0^a,W_k^b\} &=& \frac{1}{p}{f^{ab}}_c\, W_k^c
\mb{ }k=0,1,2,... \label{pbsln}\\
\{W_1^a,W_1^b\} &=& \frac{6}{p(p^2-1)} \left[
\frac{p^2-4}{5} {f^{ab}}_c\, W_2^c+
\half({d^{a}}_{cu}{f^{ub}}_d-{d^{b}}_{cu}{f^{ua}}_d)\, W_1^cW_0^d+
\right.\nonu
&&\left. \phantom{p(p^2-1)\ }
+\half{f^{a}}_{cu}{f^{b}}_{dv}{f^{uv}}_{e}\, W_0^cW_0^dW_0^e
\rule{0mm}{2.67ex}\right]
\ \ \ \ \ \ \ \ \\
\{W_1^a,W_2^b\} &=& \frac{6}{p(p^2-1)} \left[
 \frac{3(p^2-9)}{14} {f^{ab}}_c\ W_3^c+
\left(\rule{0mm}{2.67ex}\half{f^{b}}_{cu}{d^{ua}}_d
 -{f^{a}}_{cu}{d^{ub}}_d \right)
\ W_0^cW_2^d+\frac{1}{6}{f^{ab}}_u{d^{u}}_{cd}\ W_1^c W_1^d+
\right.\nonu
&&\left. \phantom{p(p^2-1)\ }+
\left(\rule{0mm}{2.67ex}\half {f^{a}}_{cu}{f^{b}}_{ev}{f^{uv}}_{d}+
\frac{1}{4} {f^{a}}_{du}{f^{b}}_{cv}{f^{uv}}_{e}+
\frac{1}{12} {f^{a}}_{eu}{f^{b}}_{cv}{f^{uv}}_{d}\right)\ 
W_0^c W_0^d W_1^e\right]
\label{pbchiant}
\end{eqnarray}
We repeat that the indices run from 0 to $n^2-1$.

Noting the identity (proved using (\ref{dd=f}) and the commutativity of
the product)
\be
{f^{a}}_{cu}{f^{b}}_{dv}{f^{uv}}_{e}\, W_0^cW_0^dW_0^e=\left(
\half{f^{ab}}_{u}{d^{u}}_{cv}{d^{v}}_{de}\, -
\frac{3}{4}({d^{a}}_{uv}{f^{ub}}_{c}-{d^{b}}_{uv}{f^{ua}}_{c}){d^{v}}_{de}
\right)\, W_0^cW_0^dW_0^e \label{fff=fdd}
\ee
and performing a change of basis
\bea
\widetilde{W}_1^a &=& \frac{p(p^2-1)}{6}W_1^a+\frac{p}{4}\,{d^a}_{bc}\, W_0^b
W_0^c
\label{newW1}\\
\widetilde{W}_2^a &=& \frac{p(p^2-1)(p^2-4)}{6}W_2^a+
5\,{d^a}_{bc}\, W_0^b
\widetilde{W}_1^c +\frac{5p(p^2-4)}{24}\,{d^a}_{bu}{d^u}_{cd}\, W_0^bW_0^cW_0^d
\ \ \ \ \ \ \ \ \ \ \ \ \ \ \label{newW2}\\
\widetilde{W}_3^a &=& \frac{p(p^2-1)(p^2-4)(p^2-9)}{6}W_3^a
\eea
we obtain the PB\footnote{We keep the notation $W_j^a$ for 
$\widetilde{W}_j^a$: throughout the text it is $\widetilde{W}_j^a$
which is used, except in the equations (\ref{pbsln}-\ref{pbchiant}) and the
convention (\ref{norm7}).}:
\begin{eqnarray}
\{W_1^a,W_1^b\} &=&\frac{1}{5}{f^{ab}}_c\, W_2^c-\frac{p^3}{16}\left(
{d^{au}}_{v}{f^{b}}_{cu}-{d^{bu}}_{v}{f^{a}}_{cu} 
\right){d^{v}}_{de}\, W_0^cW_0^dW_0^e
\label{lo1}
\end{eqnarray}
Note that in this basis, we have $W_1^0=C_2$ (i.e. $W_1^0$ is central).

Now, as the relations that we have to verify are different if $n$ is 2
or not, we specify both cases. We begin with the general case.

\subsection{The generic case $n\neq2$}

One has to verify that the PB (\ref{lo1}) obeys to the defining relations of
the Yangian.
We rewrite (\ref{yG}) as
\be
{f^{bc}}_d \{Q^a_1,Q^d_1\} +\pc{a,b,c}=
{f^a}_{qd}{f^b}_{rx}{f^c}_{sy}{f^{xyd}}
Q_0^q\,Q_0^r\,Q_0^s
\label{yGclass}
\ee
Plugging the PB
 into the left hand side of (\ref{yGclass}) leads to
\beo
\mbox{lhs} &=& -\beta_0\beta_1^2
\frac{p^3}{16}\left[\frac{1}{p}{f^{bc}}_{d}\left(
{d^{a\mu }}_{\nu }{f^{d}}_{\pi\mu}-{d^{d\mu }}_{\nu }{f^{a}}_{\pi\mu } 
\right)+\pc{a,b,c}\right]
{d^{\nu }}_{\gamma \rho }\, W_0^\pi W_0^\gamma W_0^\rho 
\eeo
This has to be compared with
\[
\mbox{rhs}\ =\ \beta_0^3{f^a}_{qx}{f^b}_{ry}{f^c}_{sz}{f^{xyz}}
\ W_0^q\, W_0^r\, W_0^s
\]
where we have used latin (resp. greek) letters for $sl(n)$ (resp. $gl(n)$) 
indices.

To prove the equality between lhs and rhs, we 
first remark, using the Jacobi identity for $f$, 
 that the index 0 can be dropped from lhs, or equivalently added to
 rhs. We choose to use $gl(n)$ indices, and come back to latin letters
 to denote them. 
\bea
\mbox{lhs} &=& -\beta_0\beta_1^2\frac{p^2}{16}{f^{ab}}_{d}\left(
{d^{cy}}_{x}{f^{d}}_{qy}-{d^{dy}}_{x}{f^{c}}_{qy} 
\right){d^{x}}_{rs}
\ W_0^q\, W_0^r\, W_0^s +\pc{a,b,c}\ \ \ \nonu
 &=&-\beta_0\beta_1^2 \frac{p^2}{8}
{f^{ab}}_{d}{d^{dy}}_{x}{f^{c}}_{yq} 
{d^{x}}_{rs}\ W_0^q\, W_0^r\, W_0^s +\pc{a,b,c}
\eea
where we have used the Jacobi identities (\ref{Jacoff}-\ref{Jacofd}). 
With (\ref{dd=f}) and the symmetry in $(q,r,s)$,
 one can
 rewrite rhs as:
\bea
\frac{1}{\beta_0^3}\mbox{rhs} &=& {f^a}_{qd}{f^b}_{rx}({d^{cx}}_y{d^{yd}}_s-
{d^{cd}}_y{d^{yx}}_s)\ W_0^q\, W_0^r\, W_0^s\nonu
&=& \left(
\half{d^{cx}}_y{f^b}_{rx}({d^{yd}}_s{f^a}_{qd}+
{d^{yd}}_q{f^a}_{sd})-\half{d^{cd}}_y{f^a}_{qd}({d^{yx}}_s{f^b}_{rx}+
{d^{yx}}_r{f^b}_{sx})\right)\ W_0^q\, W_0^r\, W_0^s\nonu
&=& \half({f^a}_{qd}{f^{by}}_{x}-
{f^{ay}}_{x}{f^b}_{qd}){d^{cd}}_y{d^{x}}_{rs}
\ W_0^q\, W_0^r\, W_0^s\nonu
&=& -\half\ {R^{abc}}_{qx}{d^{x}}_{rs}
\ W_0^q\, W_0^r\, W_0^s
\eea
Using the Jacobi identity (\ref{Jacofd}), we have
\bea
{R^{abc}}_{qx} &=& 
{f^{by}}_{x}({f^{ac}}_{d}{d^{d}}_{qy}
+{f^{a}}_{yd}{d^{cd}}_q)-(a\,\leftrightarrow\, b)\nonu
&=&({f^{ac}}_{d}{f^{by}}_{x}-
{f^{bc}}_{d}{f^{ay}}_{x}){d_{yq}}^d+({f^{by}}_{x}{f^{a}}_{yd}-
{f^{ay}}_{x}{f^{b}}_{yd}){d^{cd}}_q\nonu
&=& ({f^{ac}}_{d}{f^{by}}_{x}-
{f^{bc}}_{d}{f^{ay}}_{x}){d_{yq}}^d+
{f^{ab}}_{y}{f_{dx}}^{y}{d^{cd}}_q
\eea
Now, since rhs is invariant under cyclic permutations of $(a,b,c)$, we
can write
\bea
6\,\mbox{rhs} &=& \beta_0^3\left(2
{f^{ab}}_{d}{f^{cy}}_{x}{d_{yq}}^d-
{f^{ab}}_{y}{f_{dx}}^{y}{d^{cd}}_q+\pc{a,b,c}\right){d^{x}}_{rs}
\ W_0^q\, W_0^r\, W_0^s\nonu
&=& \beta_0^3{f^{ab}}_{d}\left(-2{f^{cy}}_{q}{d_{yx}}^{d}-
{f_{yx}}^{d}{d^{cy}}_q+\pc{a,b,c}\right){d^{x}}_{rs}
\ W_0^q\, W_0^r\, W_0^s\nonu
&=& -3\beta_0^3{f^{ab}}_{d}{f^{cy}}_{q}{d_{yx}}^{d}{d^{x}}_{rs}
\ W_0^q\, W_0^r\, W_0^s+\pc{a,b,c}
\eea
From the normalisation $\beta_0=p$, we deduce that lhs and rhs are
equal when $\beta_1^2=4$, i.e. for 
\be
Q_0^a=p\, W_0^a \ \mb{and}\ 
Q_1^a=2\, W_1^a
\ee
which ends the proof for the generic case.

\subsection{The particular case of $Y(sl(2))$}

As a normalisation, we take for the fundamental 
representation of $gl(2)$ the
matrices:
\be
t^1=\left(\begin{array}{cc} 0 & 1 \\ 1 & 0 \end{array}\right)\mb{ ; }
t^2=\left(\begin{array}{cc} 0 & -i \\ i & 0 \end{array}\right)\mb{ ; }
t^3=\left(\begin{array}{cc} 1 & 0 \\ 0 & -1 \end{array}\right)\mb{ ; }
t^0=\left(\begin{array}{cc} 1 & 0 \\ 0 & 1 \end{array}\right)
\ee
We have in that case 
\be
d^{abc}=2\delta^a_0\,\eta^{bc}+\pc{a,b,c} \mb{ and }
c_v=-8
\ee
Then, using the
special property ${f^{ij}}_{m}{f^{m}}_{kl}=-4(\eta^i_k\eta^j_l-
\eta^i_l\eta^j_k)$ valid in \underline{$sl(2)$} (i.e. when none of the
index is 0), 
we get the PB
\be
\{W_1^i,W_1^j\} = {f^{ij}}_{k}\left[\frac{1}{5}\, W_2^k
 -\frac{p^3}{2} (\vec{W_0}\cdot\vec{W_0})\, W_0^k
\right]\label{aux2}
\ee
where $\vec{x}\cdot\vec{y}=x^1y_1+x^2y_2+x^3y_3$ and the indices 
$i$, $j$, $k$ now run from 1 to 3. 

Note that for $p=2$, once the constraint 
$W_2^a=10\,[W_0^aW_1^0+\delta^a_0(\vec{W_1}\cdot\vec{W_0})]$
is applied, we recover the algebra presented in section \ref{exemple}, up to
the normalisation $J^i=W^i_0$, $S^i=2\,W^i_1$, $C_2=W_1^0$ and 
${f^{ij}}_{k}=2i{\eps^{ij}}_{k}$.

\indent

After multiplication
by ${f_{ij}}^{k}{f_{mn}}^{l}$, the relation (\ref{ysl2}) can be rewritten as:
\be
{f_{ij}}^{k} 
\{\{W_1^i,W_1^j\},W_1^l\}+{f_{ij}}^l \{\{W_1^i
,W_1^j\},W_1^k\}=32\left(
{f_{ij}}^{k}\eta^{l}_{m} +
{f_{ij}}^{l}\eta^{k}_{m}\right) W_0^mW_0^jW_1^i
\label{ysl2class}
\ee
Using the above PB and the normalisation $\beta_0=p$, we get:
\be\begin{array}{l}
\mbox{lhs}\ =\ c_v\beta_1^3\left[\frac{1}{5}\, 
\left(\rule{0mm}{2.67ex}\{W_2^k,W_1^l\}+
\{W_2^l,W_1^k\}\right)-p^2\left(\rule{0mm}{2.67ex}{f_{ij}}^{k}\eta^{l}_{m} +
{f_{ij}}^{l}\eta^{k}_{m}\right) W_0^mW_0^jW_1^i\right]\\
\mbox{rhs}\ =\ 32p^2\beta_1 \left(
{f_{ij}}^{k}\eta^{l}_{m} +
{f_{ij}}^{l}\eta^{k}_{m}\right) W_0^mW_0^jW_1^i 
\end{array}
\ee
Thus, we need to simplify the PBs (\ref{pbchiant}) in the new basis 
(\ref{newW1}-\ref{newW2}). For $sl(2)$ it
takes the form:
\begin{eqnarray}
\{W_1^i,W_2^j\} &=& {f^{ij}}_{k}\, \left[
\frac{3}{14} W_3^k+3\,W_2^0W_0^k+ 
\frac{6(3p^2-7)}{p(p^2-1)} W_1^kW_1^0
+\frac{(p^2-9)(p^2-4)}{2(p^2-1)}\ W_1^k\,{\vec{W_0}}^2 +\right. \nonu
&&\left.\phantom{{f^{ij}}_{k}\ } 
-30\,W_0^k\,(\vec{W_1}\cdot\vec{W_0})\right]
\end{eqnarray}
so that it does not contribute to lhs. Hence, we have
\be
\mbox{lhs}
\ =\ -c_v\beta_1^3\,p^2({f^{k}}_{ij}\eta^l_{m}+
{f^{k}}_{ij}\eta^l_{m})\ W_0^m W_0^j W_1^i
\ee
The relation (\ref{ysl2class}) is then satisfied for 
\be
Q^a_0=p\, W^a_0 \mb{ and }Q^a_1=2\, W^a_1
\ee
which is the same normalisation as for the generic case.

\sect{Quantum $\cW(sl(np),n.sl(p))$ algebras\label{apWq}}

In the quantum case, one has to check that the corrections to the
leading terms in the \cW-algebras do not perturb the defining 
relations of the Yangian\footnote{More exactly that the modification is the
same as the one introduced in replacing the commutative product in
$\cU(\cG^*)$ by the (symmetrised) non Abelian product of \cU(\cG).}.

\indent

{\bf -- In the general case $n\neq2$}, the calculation is quite easy. Indeed, 
the commutator takes the form
\begin{eqnarray}
{[W_1^a,W_1^b]} &=& -\frac{p^3}{16}
\left({d^{au}}_{v}{f^{b}}_{cv}-{d^{bu}}_{v}{f^{a}}_{cv} 
\right){d^{v}}_{de}\, s_3(W_0^c,W_0^d,W_0^e)+\nonu
&&+{f^{ab}}_c\, \left[ 
\frac{1}{5}\, W_2^c+\mu_1\, W_1^c +\mu_2\, {d^c}_{de}\ s_2(W_0^d,W_0^e)+
\mu_3\,  W_0^c\right]
\end{eqnarray}
where $\mu_i$ ($i=1,2,3$) are undetermined constants. However,
one remarks that the
terms containing ${f^{ab}}_c$ in $[W^a_1,W_1^b]$ do not contribute 
to (\ref{yG}). Since these are the only type of terms we add, the
calculation is identical to the classical one (up to symmetrization of
the products).

\indent

{\bf -- In the case of $\cW(sl(2p),2.sl(p))$}, we need a little more. Due to
the calculations done in the classical case, we already know that
proving (\ref{ysl2}) amounts to show that
\be
\mu_1\left(\rule{0mm}{2.1ex}
[W_1^i,W_1^j]+
[W_1^j,W_1^i]\right)+\mu_3\left(\rule{0mm}{2.1ex}[W_0^i,W_1^j]+
[W_0^j,W_1^i]\right)+\frac{1}{5}\,\left(\rule{0mm}{2.1ex}[W_2^i,W_1^j]+
[W_2^j,W_1^i]\right)\ =\ 0\label{aux1}
\ee
where the indices run from 1 to 3.
The terms corresponding to $\mu_1$ and $\mu_3$ disappear because of
the antisymmetry in $(i,j)$. Thus, we just need to compute the
corrections to the commutator $[W_2^a,W_1^b]$. Using the results of
appendix \ref{tensor}, we compute the most
general form of this commutator:
\begin{eqnarray}
{[W_1^a,W_2^b]} &=&
 {f^{ab}}_c\left[\frac{3}{14}\ W_3^c+\frac{6(3p^2-7)}{p(p^2-1)}\, s_2(W_1^c,W_1^0)
+3\,s_2(W_2^0,W_0^c)+\right.\nonu
&& \phantom{{f^{c}}_{de}\ }\left.+(
\frac{(p^2-9)(p^2-4)}{2(p^2-1)}\,\eta_{dg}\eta^c_{e}
-30\,\eta_{de}\eta^c_{g})\ 
s_3(W_0^g,W_0^d, W_1^e)\right]
+\nonu
&&+{f^{ab}}_c\ \left(\rule{0mm}{2.67ex}\nu_1\
 W_2^c +\nu_1'\ W_1^c+\nu_1''\ W_0^c\right)+\nu_2
{f^{ab}}_c\eta_{de}\ s_3(W_0^c,W_0^d,W_0^e)+\nonu
&&+\left(\nu_3\,{f^{ab}}_uf^{u}_{cd}+
\nu'_3\, \eta^{ab}\eta_{cd}+\nu''_3(\eta^a_c\eta^b_d+
\eta^a_d\eta^b_c)\right)\ W_1^cW_0^d+ \nonu
&&+
\left(\nu_4\, \eta^{ab}\eta_{cd}+\nu'_4(\eta^a_c\eta^b_d+
\eta^a_d\eta^b_c)\right)\ s_2(W_0^c,W_0^d)
\end{eqnarray}
with indices running from 0 to 3.
Looking at (\ref{aux1}), one sees that
some of the new terms 
that may appear in the right hand side of the commutator do contribute
to (\ref{aux1}). Thus, one has to check that they are not in the
true commutator. It is done thank to the Jacobi identity based on 
$(W_1^a,W_1^b,W_1^c)$ which shows (for $a,b,c$ all different) 
that\footnote{Let us note {\em en passant} that the Jacobi identity
  has just removed in the new terms those which are symmetric in $a,b$.}
$\nu_3'=\nu_3''=\nu_4=\nu_4'=0$.
We deduce that the commutator takes the form:
\begin{eqnarray}
{[W_1^a,W_2^b]} &=&  {f^{ab}}_c\left[\frac{3}{14}\ W_3^c+\frac{6(3p^2-7)}{p(p^2-1)} s_2(W_1^c,W_1^0)
+3\,s_2(W_2^0,W_0^c)+\right.\nonu
&& \phantom{{f^{c}}_{de}\ }+\left(
\frac{(p^2-9)(p^2-4)}{2(p^2-1)}\,\eta_{dg}\eta^c_{e}
-30\,\eta_{de}\eta^c_{g}\right)\ 
s_3(W_0^g,W_0^d, W_1^e)
+\nonu
&&\phantom{{f^{c}}_{de}\ }\left.+\nu_1\,W_2^c +\nu_1'\,W_1^c+
\nu_1''\,W_0^c+\nu_2\,\eta_{de}\ s_3(W_0^c,W_0^d,W_0^e)
+\nu_3\,{f^{c}}_{de}W_1^dW_0^e
\rule{0mm}{2.67ex}\right]\nonumber
\end{eqnarray}
so that (\ref{aux1}) and hence (\ref{ysl2}) are satisfied.

\sect{Tensor products of some finite dimensional representations
  of $sl(n)$\label{tensor}}

We want here to compute the tensor product of the \cG-adjoint
  representation by itself several times, 
for $\cG=sl(n)$. 
We will also need to select the totally symmetric part of these
  products.

{F}or such a purpose, we use Young diagrams, which allow us to determine
the decompositions:
\[
\cG\otimes\cG= (2,0,..,0,2)\oplus 2\,(1,0,..,0,1)\oplus 
(2,0,..,0,1,0)\oplus (0,1,0,..,0,2)\oplus (0,1,0,..,0,1,0)\oplus 
(0,..,0)
\]
where we have denoted by $\cG=(1,0,\dots,0,1)$ the adjoint
representation.

It remains to select the (anti-)symmetric part of these products. For 
$\cG\otimes\cG$, the calculation has already been done (see
e.g. \cite{gourdin}) and reads:
\bea
&& S_2(\cG) = (\cG\otimes\cG)_{sym}= (2,0,..,0,2)\oplus
(1,0,..,0,1)\oplus (0,1,0,..,0,1,0)\oplus (0,..,0)
\ \ \ \ \ \ \ \ \\
&& \Lambda_2(\cG) = (\cG\otimes\cG)_{skew}=(1,0,..,0,1)\oplus 
(2,0,..,0,1,0)\oplus (0,1,0,..,0,2)
\eea
As far as $S_3(\cG)$ is concerned, we already now that this sum of
representations belongs to $(S_2(\cG)\otimes\cG)_{sym}$, which
decomposes as
\bea
(S_2(\cG)\otimes\cG)_{sym} &=& (3,0,\dots,0,3)\oplus 3\,(2,0,\dots,0,2)\oplus 
3\,(1,0,\dots,0,1)\oplus 3\,(0,1,0,\dots,0,1,0)\oplus \nonu
&& \oplus\, 2\,(1,1,0,\dots,0,1,1)\oplus 
2\,[(2,0,\dots,0,1,0)\oplus (0,1,0,\dots,0,2)]\oplus \nonu
&& \oplus\, (0,0,1,0,\dots,0,1,0,0)\oplus 
[(3,0,\dots,0,1,1)\oplus (1,1,0,\dots,0,3)]\oplus \nonu
&& \oplus \,
[(1,1,0,\dots,0,1,0,0)\oplus (0,0,1,0,\dots,0,1,1)]\oplus (0,\dots,0)
\label{s2G}
\eea
This implies that we must have
\bea
S_3(\cG) &=& a\,(3,0,\dots,0,3)\oplus b\,(2,0,\dots,0,2)\oplus 
c\,(1,0,\dots,0,1)\oplus d\,(0,1,0,\dots,0,1,0)\oplus \nonu
&& \oplus \,
e\,(1,1,0,\dots,0,1,1)\oplus f\,(0,0,1,0,\dots,0,1,0,0)\oplus 
 m\,(0,\dots,0)\oplus \nonu
&& \oplus\, g\,[(2,0,\dots,0,1,0)\oplus (0,1,0,\dots,0,2)]\oplus \nonu
&& \oplus \,
h\,[(3,0,\dots,0,1,1)\oplus (1,1,0,\dots,0,3)]\oplus \nonu
&& \oplus \,
i\,[(1,1,0,\dots,0,1,0,0)\oplus (0,0,1,0,\dots,0,1,1)]
\label{s3}
\eea
with each multiplicity in (\ref{s3}) lower or equal to the corresponding
multiplicity in (\ref{s2G}). But we know the dimension of $S_3(\cG)$:
it is the dimension of a totally symmetric tensor with 3 indices in a
space of dimension dim\cG=$n^2-1$, i.e. 
$\frac{n^2(n^4-1)}{6}$. Computing
this dimension with (\ref{s3}) leads to only two possible solutions for the
parameters: $a=e=f=1$, $h=i=0$, $b=d=m$, $c=3-m$ and $g=2-m$ with $m=0$ or 1. 
As $m$ is the multiplicity of the trivial
representation in $S_3(\cG)$, we deduce that, for\footnote{The case
  $\cG=sl(2)$ is treated below.} $\cG=sl(n),\
n\neq2$, we have
 $m=1$ (since $d_{abc}$ belongs to this space). Thus
\bea
S_3(\cG) &=& (3,0,\dots,0,3) \oplus (2,0,\dots,0,2)\oplus 
2\,(1,0,\dots,0,1)\oplus (0,1,0,\dots,0,1,0)\oplus \nonu
&& \oplus \,
[(1,1,0,\dots,0,1,1)\oplus (0,0,1,0,\dots,0,1,0,0)]\oplus \nonu
&&\oplus\, [(2,0,\dots,0,1,0) \oplus (0,1,0,\dots,0,2)]\oplus 
 (0,\dots,0)
\eea

{F}inally, the multiplicity of the trivial representation in the tensor
products occuring in section \ref{YWq} is computed through the remark
that, in $sl(n)$, the tensor product of two finite dimensional
irreducible representations $R$ and $R'$ contains the
trivial representation if and only if $R$ and $R'$ are conjugate. 
 In that case, the multiplicity is 1. This
leads to the multiplicities given in (\ref{mult}). We give also a
basis for the corresponding spaces. To be complete, let us mention the
bases:
\begin{equation}
\begin{array}{ll}
\Lambda_2(\cG)\otimes \Lambda_2(\cG)\ :\ &
\left\{\begin{array}{l}
t^{ab}_{cd} = {f^{ab}}_{e}{f^{e}}_{cd}\\
t^{ab}_{cd} = {d^{a}}_{ce}{d^{eb}}_{d}-{d^{b}}_{ce}{d^{ea}}_{d}\\
t^{ab}_{cd} = {f^{a}}_{ce}{d^{eb}}_{d}-{f^{b}}_{ce}{d^{ea}}_{d}
\end{array}\right.  \\ \\
\Lambda_2(\cG)\otimes S_3(\cG)\ :\ &
\left\{\begin{array}{l}
t^{ab}_{cde} = {f^{ab}}_{c}\eta_{cd}+ \pc{c,d,e}\\
t^{ab}_{cde} = {f^{ab}}_{g}{d^{g}}_{cm}{d^{m}}_{de}+ \pc{c,d,e}\\
t^{ab}_{cde} = (\eta^{a}_{c}{d^{b}}_{de}-\eta^{b}_{c}{d^{a}}_{de})
+ \pc{c,d,e}\\
t^{ab}_{cde} = ({f^{a}}_{cg}{d^{gb}}_{m}-{f^{b}}_{cg}{d^{ga}}_{m}){d^{m}}_{de}
+ \pc{c,d,e}
\end{array}\right. 
\end{array}
\end{equation}

In the case of $sl(2)$, we need more informations. Fortunately, the
calculation is easier in that case, and we can go further. Indeed, we
have (with $\cD_j$ the $(2j+1)$-dimensional representation of $sl(2)$):
\be
(\cD_1\times\cD_1)_{sym}=\cD_0\oplus\cD_2\ ;\ 
(\cD_1\times\cD_1)_{skew}=\cD_1\ ;\ 
S_3(\cD_1)=\cD_1\oplus\cD_3
\ee
which leads to the multiplicities and tensors:
\be
\begin{array}{lll}
M_0[(\cD_1\times\cD_1)_{skew}\times(\cD_1\times\cD_1)_{skew}]=1 &:&
{f^{ab}}_u{f^{u}}_{cd}\sim\eta^a_c \eta^b_d-\eta^a_d \eta^b_c\\
&& \\
M_0[(\cD_1\times\cD_1)_{sym}\times(\cD_1\times\cD_1)_{sym}]=2 &:&
\left\{\begin{array}{l}{\eta^{ab}}{\eta}_{cd} \\
\eta^a_c \eta^b_d+\eta^a_d \eta^b_c
\end{array}\right.\\
&& \\
M_0[(\cD_1\times\cD_1)_{skew}\times(\cD_1\times\cD_1)_{sym}]=0 && - \\
&& \\
M_0[(\cD_1\times\cD_1)_{skew}\times S_3(\cD_1)]=1 & :&
{f^{ab}}_c{\eta}_{de}+\pc{c,d,e}\\
&& \\
M_0[(\cD_1\times\cD_1)_{sym}\times S_3(\cD_1)]=0 && -
\end{array}
\ee


\end{document}